# Monitoring of Wild Pseudomonas Biofilm Strain Conditions Using Statistical Characterisation of Scanning Electron Microscopy Images

Suparna Dutta Sinha[1], Saptarshi Das[2, 3*], Sujata Tarafdar[1], and Tapati Dutta[3]

1) *Condensed Matter Physics Research Centre, Department of Physics, Jadavpur University, Kolkata 700032, India*

2) *Department of Physics, University of Cambridge, JJ Thomson Avenue, Cambridge CB3 0HE, UK*

3) *Department of Power Engineering, Jadavpur University, Salt Lake Campus, LB-8, Sector 3, Kolkata-700098, India*

4) *Physics Department, St. Xavier's College, Kolkata 700016, India*

Author's Emails:

sduttasinha@gmail.com (S. Dutta Sinha)

saptarshi@pe.jusl.ac.in, sd731@cam.ac.uk (S. Das*)

sujata@phys.jdvu.ac.in, sujata_tarafdar@hotmail.com (S. Tarafdar)

tapati_mithu@yahoo.com (T. Dutta)

**Abstract:**

The present paper proposes a novel method of quantification of the variation in biofilm architecture, in correlation with the alteration of growth conditions that include, variations of *substrate* and *conditioning layer*. The polymeric biomaterial serving as *substrates* are widely used in implants and indwelling medical devices, while the plasma proteins serve as the *conditioning layer*. The present method uses descriptive statistics of FESEM images of biofilms obtained during a variety of growth conditions. We aim to explore here the texture and fractal analysis techniques, to identify the most discriminatory features which are capable of predicting the difference in biofilm growth conditions. We initially extract some statistical features of biofilm images on bare polymer surfaces, followed by those on the same substrates adsorbed with two different types of plasma proteins, *viz*. Bovine serum albumin (BSA) and Fibronectin (FN), for two different adsorption times. The present analysis has the potential to act as a futuristic technology for developing a computerized monitoring system in hospitals with automated image analysis and feature extraction, which may be used to predict the growth profile of an emerging biofilm on surgical implants or similar medical applications.

## 1. Introduction

Biofilms are ubiquitous in nature [1–3], industrial locations [4–6], surgical infections, chronic wounds [7–9] and other physiological conditions, where bacteria abandon their planktonic status and choose to survive in microbial communities associated with a surface [10–12]. In biofilms, the bacteria remain encased within a self-secreted exopolysaccharide matrix [13–15]. Biofilm formation has been recognized as the primary virulence factor in peri-operative and post-operative infections as well as in a variety of chronic infections [16,17]. Bacteria in biofilms usually present higher resistance to antibiotics [18–20] and higher tolerance to the immune system [21,22] compared to its planktonic counterparts, posing the biggest challenge to public healthcare. Hence understanding the emergence and growth patterns of biofilms through continuous nondestructive monitoring [23,24] is the fundamental requirement for controlling biofilm growth.

Biofilm infections within a physiological environment can grossly have either tissue-related or device-related origin. The diversity in the conditions leading to the development of such biofilms, result in the difference in their physical and chemical characteristics - making biofilm monitoring methods extremely application specific. While microbiologically influenced corrosion (MIC) [25,26] represents the main cause of degradation of implanted devices of metallic origin, implants from polymer biomaterials form a robust and versatile class having unparalleled durability, biocompatibility [27], hemocompatibility [28,29], anti-thrombogenicity [30] and resistance to degradation and calcification [31]. The present paper focuses on a computerized image monitoring system for device-related biofilms on widely used polymer biomaterials, which will enable a much better understanding of the processes leading to the emergence of biofilms and aid in the quantification of biofilm architecture on different substrates in a variety of growth conditions.

### 1.1. Fundamentals of Bacterial Growth and Biofilm Formation

Bacterial biofilms are three-dimensional sessile structures consisting of bacteria encapsulated within hydrated extracellular polymeric substances (EPS) on a substrate [32]. The exopolysaccharide matrix facilitates irreversible attachment between the biofilm cells and the substrate while simultaneously maintaining intercellular interactions, giving rise to a biofilm architecture specific to a particular substrate and definite bacterial strain [33]. Medical implants from polymer biomaterials, often serve as the nidus to which bacteria can irreversibly attach via hydrophobic or electrostatic attractions during a surgical insertion,and proliferate in the form of biofilms.

Biofilm-producing bacteria usually enter the body during the process of implantation, or exist on the surface of the implant pre-surgery and colonize the implanted device. Advances in infection control strategies include improved operating room, ventilation, sterilization methods, barriers, surgical techniques, and availability of antimicrobial prophylaxis. Despite these activities, surgical site infections (SSI) remain a prime cause of morbidity and mortality among hospitalized patients. This may be partly explained by theemergence of antimicrobial-resistant pathogens [34,35], mostly residing in biofilms, increased numbers of immuno-compromised elderly surgical patients and execution of increased numbers of prosthetic implant operations. Microbial contamination of the surgical site is however a necessary precursor of surgical site infections. Microorganisms may contain or produce toxins and other substances that increase their ability to invade a host, produce damage within the host, or survive on, or in host tissue. Gram-negative bacteria such as *Pseudomonas aeruginosa* after gaining access to the surgical site, produce a "slime" (mentioned as EPS), which physically shield the bacteria from phagocytes and inhibits the binding or penetration of antimicrobial agents [36].

Many sets of stringent preventive measures have been deployed with the aim of preventing infections at the surgery site. However based on reports from National Nosocomial Infections Surveillance (NNIS) system, which monitors trends in nosocomial infections in U.S., acute-care hospitals surgical site complications, mostly accruing from biofilm-related infections, are the third most frequently reported nosocomial infection accounting for 14% to 16% of all nosocomial infections among hospitalized patients [37]. Hence to reduce the risk of biofilm related infections, apart from the sets of clinical preventive measures, a systematic but realistic interdisciplinary approach must be applied with the awareness that this risk is influenced by characteristics of the patient, nature of implanted device, mode of surgery and maintenance of sterilization of personnel and hospital.

A host of non-destructive methods have been employed to qualitatively and quantitatively study the underlying processes that govern the morphological development of biofilms with respect to different surfaces. These include microscopic techniques, image analysis, spectrochemical methods [24] including Fourier transform-infrared spectroscopy [38], electrochemical [39], calori-metric [40] and piezoelectric [41] approaches. However, rarely any effort has been applied to analyse the architecture of emerging biofilms from a hospital environment that can be of prime utility to healthcare professionals in administering the requisite antibiotics at the earliest opportunity. The computerized monitoring system which forms the topic of the present paper is a significant step towards the development of a futuristic methodology through which a non-invasive online monitoring of the surface of an implanted device may be made possible. The primary aim of the research is to detect the early emergence and growth pattern of biofilm on an implant surface, should there be one developing at the surgery site, through existing image analysis techniques. The present method of monitoring is a unique non-destructive approach to this end. We analyse and quantify biofilm architectures through image analysis of FESEM data, using existing image processing techniques. This technique may be conveniently availed by health professionals in a hospital environment through a computer monitor, only by suitable coupling of a laparoscopic arrangement and FESEM at the surgery site. The microscopic arrangement will produce regular, clear, non-distorted images of the implant surface, with spatial resolution down to almost 1 nanometre, during the post-operative stages of an implant surgery. Early detection of biofilm growth, perceived through image analysis techniques, instead of invasive methods or use of colorimetric assays (which may be toxic to the physiological environment), will enable timely administration of antibiotic prophylaxis, before the infecting microbes turn antibiotic resistant- all eviating the morbidity and mortality rates in biofilm-related implant infections by a considerable amount.

*1.2. Brief Survey of the Existing Biofilm Image Analysis Techniques*

This section, briefly reviews the contributions previously made in existing literature on biofilm quantification using various imaging techniques through different physical or chemical features, as well as previous works on mathematical modelling of biofilm growth. In contemporary research, several methods of biofilm characterisation using statistical techniques has been reported on 2D SEM images [42–48]. Yang *et al.* [49] and Jackson *et al.* [50] have studied several textural features of microscope images, e.g. entropy, angular second moment, inverse difference moment, as well as morphological features, e.g. fractal dimension, porosity, run length, diffusion distance etc. Several image thresholding methods based on various entropy measures e.g. local entropy, joint entropy, relative entropy, Renyi's entropy have been explored by Yang *et al.* [51]. A software tool for biofilm image segmentation, intensity quantification and spatial arrangement analysis has been reported in Daims *et al.* [52]. New features in 2D microscope images have also been discussed to characterise the biofilm structure e.g. characteristic length by Milferstedt *et al.* [53], and diameter of cones created on the surface by Perni and Prokopovich [54].

There have been similar analysis of biofilm images using 3D confocal laser scanning microscopy (CLSM) by Beyenal *et al.* [55], using various textural features (energy, entropy, homogeneity), as well as volumetric parameters (run length, aspect ratio, diffusion distance, fractal dimension). Similar confocal image analyses have been carried out by Mueller *et al.* [56], Truong *et al.* [57] and Ngo *et al.* [58], using features like biovolume [59], area to volume ratio, thickness, roughness, horizontal/vertical spreading etc. Some additional features like area distribution, surface to volume ratio have also been studied on confocal images, by Heydorn *et al.* [60, 61], Sandal *et al.* [62], Herzberg and Elimelech [63], using a software called Comstat. Image thresholding applied to confocal images has been described by Xavier *et al.* [64] in order to quantify biovolume and interfacial area. Studies of porosity estimation using confocal images have been performed by Lewandowski [65]. Based on biovolume, thickness and roughness measure of confocal images, six different species of bacteria have been distinguished in Bridier *et al.* [66]. Image segmentation of 2D slices of confocal images have been presented by Kyan *et al.* [67] while image segmentation of the 3D confocal images by Yerly *et al.* [68]. Geostatistical analysis has been carried out on the confocal images by Kim *et al.* [69].

The surface roughness of biofilms have characterized using both height and phase images by Auerbach *et al.* [70] using atomic force microscope (AFM) images. Phase contrast microscopy based features have been used by Liu *et al.* [71] to classify bacteria morphotypes e.g. roundness, elongation, compactness, maximum curvature, diameter, width, length, width/length ratio, area etc. Image thresholding and image registration related issues for AFM and CLSM images have been discussed by Webb *et al.* [72]. Bacterial adhesion has been explored in [73] using AFM images through quantification of several surface characteristics e.g. thickness, roughness, advancing and receding contact angle, contact angle hysteresis etc.

There have been several studies on biofilm characterisation involving multiple imaging techniques. Surman et al. [74] have compared different microscope imaging techniques e.g. SEM, AFM, CLSM, transmission electron microscopy (TEM), environmental SEM (ESEM), episcopic differential interference contrast microscopy (DIC) with and without fluorescence, Hoffman modulation contrast microscopy (HMC) etc. Similar comparison have been done between SEM and TEM by Sangetha et al. [75] and Espinal et al. [76], between SEM and confocal imaging by Villena et al. [77] and Ng et al. [78] and between SEM and variable pressure SEM (VPSEM) Weber et al. [79]. A comparison between SEM, CLSM, and phase contrast microscopy has been reported in Norton et al. [80]. Bacterial adhesion have also been explored using multiple imaging techniques including SEM, AFM, X-ray photoelectron spectroscopy (XPS), TEM and ultraviolet spectroscopy by Liu et al. [81]. Bacteria removal performance has been compared by DeQueiroz and Day [82], using different imaging techniques like SEM, TEM, CSLM, and analysed using Fourier transform infrared (FTIR) spectroscopy. Identical analysis has also been used in other applications like material degradation [83]. Comparison of biofilm characterization has been done through combination of SEM, CLSM and light microscopy using gram staining by Kania et al. [84]. It is revealed that often combined approaches e.g. SEM and CLSM give better characterisation rather than using a single imaging technique, as reported in [85].

A first principle partial differential equation (PDE) based mathematical modelling for bacterial growth has been established in Kreft et al. [86] where the simulated images are quantified using the above mentioned features along with heterogeneity and contrast. A different 2D mathematical model using cellular automata theory has been used to quantify the biofilm inner porosity by Hermanowicz [87]. Apart from the grayscale images, the usefulness of various 2D/3D coloured image quantification softwares has been reviewed by Daims and Wagner [88].

*1.3. Present Work and Contributions*

In the present work, we focus on the analysis of field emission scanning electron microscopy (FESEM) image analysis, due to its simplicity and non-destructive recording procedure. We show here that by using a combination of popular image textural features, it is possible to discriminate between several conditions of growth of the biofilm – as presented in the form of three distinct hypotheses. We also report their statistical significance level, using our experimental data. The statistical hypothesis tests are formed to find out whether there are any fundamental difference in the statistical characteristics of the biofilms grown using different substrates and adsorbed protein. Our prime aim is to find out the most significant features to support these hypotheses which are capable of indicating the difference in biofilm growth conditions from an image feature based pattern analysis perspective.

2. **Materials and Methods**
*2.1. Biomaterials and Production of the Conditioning Layer*

Commercially available clinical grade High density polyethylene (HDPE) and poly-tetra-fluoro-ethylene (PTFE) used in orthopaedic implants and venous catheters respectively, were obtained after fine machine polishing in square configuration (10mm × 10mm) from Plastic Abhiyanta Ltd, India. These samples of biomaterials have been referred to as polymer chips throughout this article.

After polishing, the samples were cleaned by 2 min ultrasonication in a 35 kHz ultrasonic bath (Rivotek Instruments, India) and thoroughly rinsed with demineralized water. The water used in all our experiments was of HPLC grade (Lichrosolv) from Merck India. Tris EDTA buffer solution pH 7.4 was obtained from Sigma Aldrich, USA. Bovine serum albumin (Fraction V) and Fibronectin from human plasma (CAS Number 86088-83-7, MDL Number MFCD00131062) lyophilized powder, MW: 45 kDa were obtained from M P Biomedicals, USA. PTFE chips were autoclaved, while HDPE chips being non-autoclavable,were rinsed twice with ethanol, blow dried and preserved in a vacuum desiccator. BSA solution ofconcentration1.5 mg/ml, prepared with buffer solution of pH 7.4 and left for about a week with intermittent mixing to dissolve the BSA completely, was used for adsorption purposes. The reconstituted BSA solution taken in 10 ml glass vials, each containing a single polymer chip of either HDPE or PTFE were kept for 9 hours, and 24 hours. The adsorption times are referred here as *exposure time* '$\tau$', where $\tau^{max}$ and $\tau^{min}$ were 24 hours and 9 hours, respectively. After the stipulated time, the chips were removed from the BSA solutions, rinsed with water, and finally blow dried and preserved in a desiccator ready for growing biofilms. The chips obtained from adsorption experiments possessed different degrees of BSA adsorbed on them, and have been termed as the *conditioning layer*. These could not be sterilized further as it would lead to the denaturation of the adsorbed protein.

The above process of adsorption was repeated for HDPE and PTFE chips with a reconstituted Fibronectin (FN) solution of concentration 1 μg/l. After 24 hours, all chips with *conditioning layer* of Fibronectin were removed from the Fibronectin solution and dipped thrice in demineralized water. The exposure time τ in Fibronectin is kept as 24 hours. The rest of the chips were left untreated to serve as controls. The concentrations of reconstituted BSA and Fn solutions werekept in proportion with their concentrations inthe plasma. The chips were all treated at 30°C, and τ was kept sufficiently high in all the cases, to obtain a complete surface coverage. After the production of the *conditioning layer*, the treated chips were thoroughly rinsed in Phosphate Buffer Saline (PBS) in order to remove any non-adsorbing protein molecules, finally leaving only irreversibly adsorbed protein molecules on the polymer surfaces. The untreated chips were rinsed with demineralized water and PBS, and all the experimental chips were finally blow-dried and preserved in a vacuum desiccator.

### 2.2. Bacterial Strain and Culture Condition

A wild type strain of *Pseudomonas aeruginosa* obtained and isolated from uro-catheters of patients having urinary tract infections (UTIs) at the Department of Urology, Institute of Post Graduate Medical Education & Research, Kolkata, India was used in this study. The strain was previously reported to be a strong biofilm former by 96 well micro-titer plate assay. The goal was to prove experimentally that the pathogenic strain responsible for urinary infections (forming biofilm on silicon rubber) [89] is very much liable to affect orthopaedic implants.

Each type of frozen culture was initially inoculated on Tripticase soy agar and incubated at 37°C for 24 hours. Each culture was then transferred via swab to a buffer solution and a suspension equivalent to a Macfarland 0.5 (~1.5 x 10⁸ CFU/ml) was prepared. This suspension was diluted at 1:100 and 1 ml was used to inoculate 100 ml of sterile LB broth (Luria Bertanii Agar (LBB) obtained from Himedia, India).The bacteria grown overnight in LBB at 37°C were diluted in the same medium to an optical density of 0.5 at 600 nm and ready for use as bacterial culture for growing biofilms.

### 2.3. Biofilm Growth Condition

The diluted culture obtained (in section 2.2), was poured over the surface of the treated chips (from section 2.1) placed in the wells of 24 well tissue culture plates (Tarsons, India) and on untreated chips kept in a separate 24 well plate, at 25°C. Each 24 well plate contained a separate set of chips with a definite adsorption time, and each system was closed and sealed without addition or removal of any component with the exception of broth. The sterile LBB was added carefully from time to time to avoid desiccation and incubated at 37°C for 7 days with shaking at 180 rpm. Each set of experiment was performed in triplicate. The plates were sealed and placed on the shaker plate of the biological oxygen demand (BOD) incubator. Care was taken to ensure that each plate was in an upright position during rotation, without tilting, which might affect the growth condition of the biofilms.

After the entire 7-day growth period, the polymer chips were aseptically removed and washed thrice with phosphate buffered saline (PBS pH 7.2). This step eliminated all the free floating bacteria and only the sessile forms remained attached to the surface. The chips were then air dried and prepared for FESEM measurements. The chips with attached bacterial cells were covered with 2.5% glutaraldehyde and kept for 3 hours then passed once through the graded series of 25%, 50% and 75% ethanol, and twice through 100% ethanol, for ten minutes each. These are finally transferred to the critical point drier and kept overnight to make them ready for biofilm analysis.

To compare the architecture of the biofilms produced by the clinically isolated strain of *Pseudomonas aeruginosa* on different substrates after 7 days, FESEM measurements were conducted at 5.0 kV-10 kV in a field emission scanning electron microscope (FESEM: Inspect F50, FEI Europe BV, The Netherlands; FP 2031/12, SE Detector R580). For this purpose the dried polymer chips, with and without biofilms, were sputter-coated with a 3-nm thick conductive layer of gold.

### *2.4. Experimental Conditions and the Research Question*

Our experiments reveal a difference in biofilm architecture of a specific strain of bacteria, in response to variations in the nature of *substrate* and *conditioning layer*. We report here four different combinations of growth condition of biofilms. The main motive is to identify the most significant features that may be used to distinguish between biofilms grown on different *substrates,* and having dissimilar *conditioning layers*. We have studied four different cases and compared biofilms grown on bare HDPE/PTFE, with that grown on HDPE/PTFE surfaces adsorbed with BSA/Fibronectin for 9 hours and 24 hours. Three different image analysis techniques have been adopted for statistical feature extraction and hypothesis testing [90]:

- Image gray level histogram (first order statistics) – central moments and standardized moments
- Image texture analysis (second order statistics) based on relative position of the pixels using gray level co-occurrence matrix (GLCM)
- Fractal analysis and Haussdorf dimension

We hypothesize that, if there is any significant difference between the biofilms determined by differences in –

1) adsorption time of the proteins

2) nature of protein used as conditioning layer and

3) nature of substrate used for growing the biofilms

it is possible to discriminate between these conditions, using some textural features of the SEM images. This essentially helps in the statistical characterization of the biofilms when their experimental conditions are not known in a real implant monitoring scenario. The analyses was carried out on images with a fixed magnification factor of 10,000x.

Here we make the following three hypotheses and also test the statistical significance levels of each feature in comparing these hypotheses, as follows:

1) Comparison with respect to adsorption time: Biofilms grown on HDPE control (bare surface), HDPE adsorbed with BSA for 9 hours and 24 hours

2) Comparison with respect to conditioning layer: Biofilms grown on HDPE control (bare surface), HDPE BSA 24 hours, HDPE FN 24 hours

3) Comparison with respect to substrate: Biofilms grown on another (PTFE) substrate - PTFE control (bare surface), PTFE BSA 24 hours, PTFE FN 24 hours.

We use here, the one way analysis of variance (ANOVA) to test the hypothesis that the population mean in each group are different along with the non-parametric version of ANOVA test and we also report the statistical significance of the best features that are found capable of discriminating between these groups.

### 3. Description of the Statistical Features of Biofilm Images
#### 3.1. Feature from First Order Statistics of Images

The first order statistics of grayscale images refer to the statistical parameters extracted from the one dimensional probability density function of the pixel intensities. The grayscale biofilm image is normally considered as a 2D matrix of random intensity $i$ with a histogram bin count $p_i$ expressed as a ratio between the number of pixels with an intensity $i$ and the total number of pixels $N$. The maximum number of graylevel is given by $N_g$, as per standard grayscale image processing whose value is 256.

$$p_i = (\text{pixels with graylevel } i)/N \qquad (1)$$

From the histogram count $p_i$, the central moments (mean subtracted) and standardized moments (mean removed, followed by unit standard deviation) can be calculated. The four descriptive statistics (mean, standard deviation, skewness, kurtosis) can now be calculated directly from the normalized image histograms as follows:

$$\mu = \mathbb{E}[I] = \sum_{i=0}^{N_g-1} i p_i, \quad \sigma = \sqrt{\mathbb{E}[I-\mu]^2} = \sqrt{\sum_{i=0}^{N_g-1} (i-\mu)^2 p_i},$$

$$\gamma = \mathbb{E}\left[(I-\mu)/\sigma\right]^3 = \sum_{i=0}^{N_g-1} (i-\mu)^3 p_i \Big/ \sigma^3, \qquad (2)$$

$$\beta = \mathbb{E}\left[(I-\mu)/\sigma\right]^4 = \sum_{i=0}^{N_g-1} (i-\mu)^4 p_i \Big/ \sigma^4.$$

Here $\mathbb{E}$ represents the mathematical expectation operator.

Similarly the energy ($E$) and entropy ($H$) of the pixel intensity can also be calculated from the histogram as:

$$E = \sum_i p_i^2, \quad H = -\sum_i p_i \log_2(p_i) \qquad (3)$$

The smoothness texture (SM) [91] is calculated from the variance ($\sigma^2$) as:

$$SM = 1 - \left(1/(1+\sigma^2)\right) \qquad (4)$$

Example of histograms for grayscaled biofilm SEM images on different cases of HDPE substrate have been shown in Figure 1, with 256 discrete intensity levels. It is evident from Figure 1 that, while for the bare HDPE surface the histogram appears continuous, in the other cases few discrete pixel intensities are dominant and the histograms appear sparse. Here all the image histograms have been reported on a normalized scale with respect to the total bin count so that the area under the histograms becomes unity, and hence they could be interpreted as probability density functions (PDFs) in pixel intensity. A similar analysis of the histograms of the biofilms formed on the PTFE surface in Figure 2, shows a more continuous distribution of pixel intensities for all the cases, with a more left-skewed distribution, compared to those in the HDPE case as shown in Figure 1. It is important to note here that the histograms shown in this section do not include all the image dataset but explore a few example images, as representative examples for the feature extraction from grayscale images.

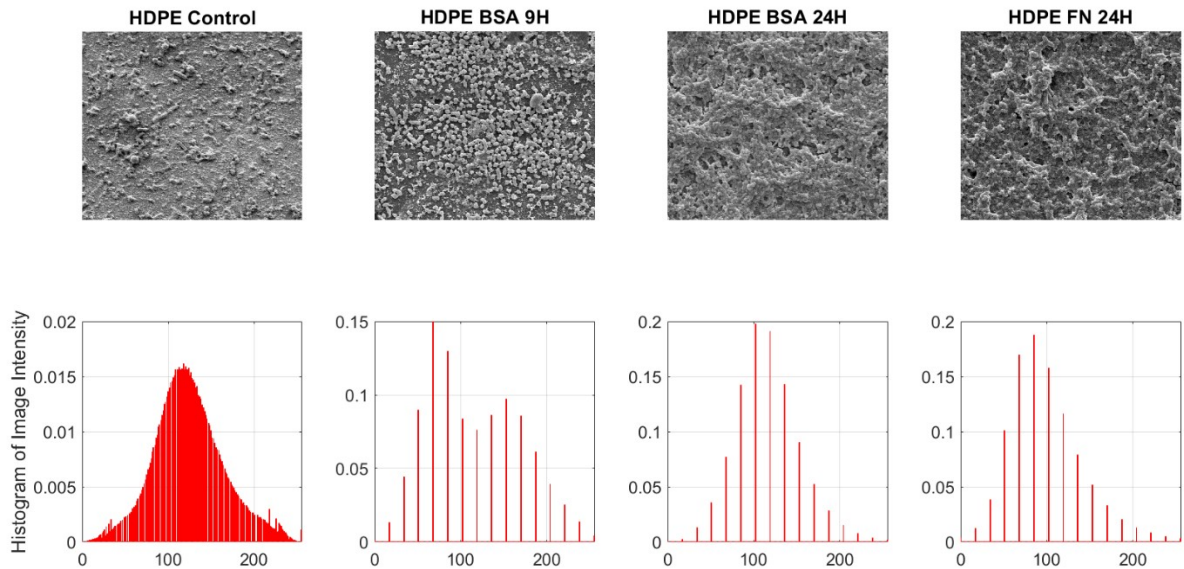

*Figure 1: (Top panel) Grayscale biofilm images for HDPE, (Bottom panel) Normalised histogram of the pixel intensities.*

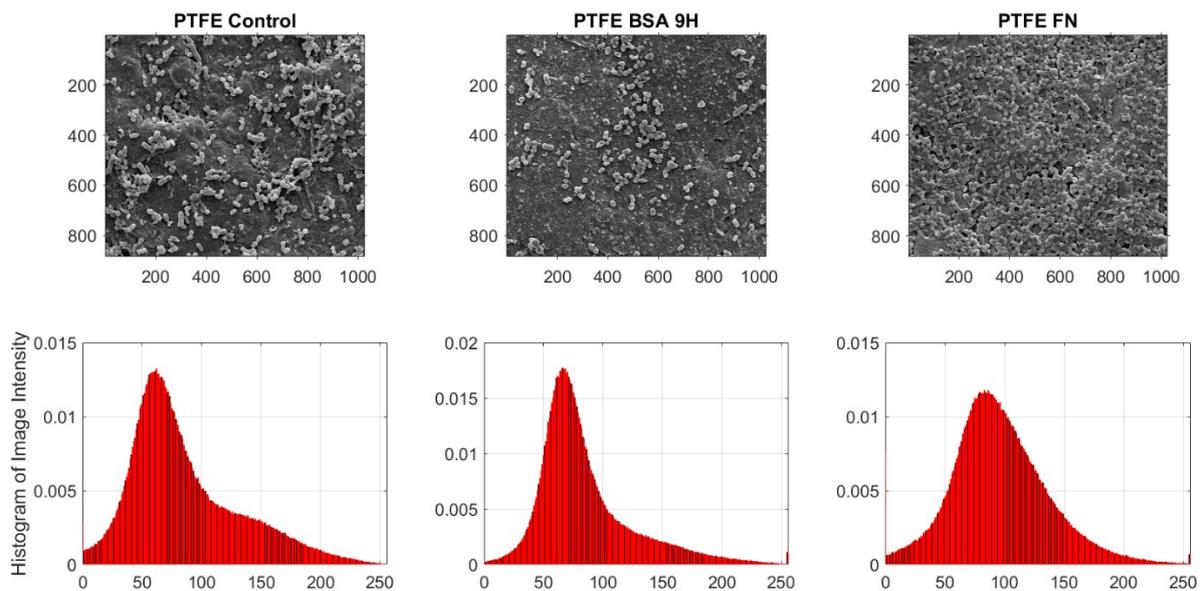

*Figure 2: (Top panel) Grayscale biofilm images for PTFE, (bottom panel) Normalised histogram of the pixel intensities.*

### 3.2. Features from Second Order Statistics of Images

The second order statistics of grayscale images refer to the spatial correlation analysis on pixel intensities along both the *x* and *y*-direction. The image is first converted to Gray Level Co-occurrence Matrix (GLCM), denoted as $p_{i,j}$ which measures the two dimensional relative frequencies, or the probability of two pixels with intensity *i* and intensity *j* occurring in a neighbourhood of distance $(\Delta x, \Delta y)$ [92]. The GLCM matrix can either be interpreted as co-occurrence of two intensities in the image separated by pixel distance $(\Delta x, \Delta y)$ as $p(i,j|\Delta x, \Delta y)$ or, separated by radial distance *d* and angle $\theta$ as $p(i,j|d,\theta)$. Depending on the discrete graylevels ($N_g$) of the image as described in earlier section, the GLCM $p(i,j)$ is derived as a $G \times G$ matrix, for different combinations of either $(\Delta x, \Delta y)$ or $(d, \theta)$. Different features can now be extracted from the GLCM matrix [93,94], as follows:

$$\text{contrast} = \text{variance} = \text{inertia} = \sum_{i,j} |i-j|^2 p(i,j) \tag{5}$$

$$\text{correlation} = \sum_{i,j} \frac{(i-\mu_x)(j-\mu_y) p(i,j)}{\sigma_x \sigma_y} \tag{6}$$

$$\text{energy} = \text{uniformity} = \text{angular second moment} = \sum_{i,j} (p(i,j))^2 \tag{7}$$

$$\text{homogeneity} = \text{inverse difference moment} = \sum_{i,j} \frac{p(i,j)}{1+|i-j|} \tag{8}$$

Here, $\sum_{ij}(.) = \sum_{i=0}^{N_g-1} \sum_{j=0}^{N_g-1} (.)$ and $\{p_x(i), p_y(j)\}$ are the marginal probability densities, summed over the $y$ and $x$-direction respectively, i.e.

$$p_x(i) = \sum_{j=0}^{N_g-1} p(i,j), \quad p_y(j) = \sum_{i=0}^{N_g-1} p(i,j) \tag{9}$$

Parameters $\{\mu_x, \mu_y, \sigma_x, \sigma_y\}$ denote the mean and standard deviation of the marginal densities along the two axis in (9) respectively. For a homogeneous image there will only be few high amplitude elements in $p_{i,j}$, so the 2D energy or ASM in equation (7) will be higher. The contrast in equation (5) measures local intensity variation and favours contributions between elements away from the diagonal of GLCM. Also, because of the inverse relationship in the homogeneity estimate in equation (8), the inhomogeneous areas of the GLCM contribute less, whereas homogenous areas makes it a higher value. The correlation in equation (6) measures the linear dependence between image pixels relative to each other in both the directions.

These features could have been extracted from the raw images or after different transformations to see if the discriminatory capability of the features is enhanced under such pre-processing. Here we tested on both the raw grayscale images and its contrast enhanced version via histogram equalization (to get close to a uniform distribution of the pixel intensity) and it was discovered that the raw images are more reliable to extract features because automated preprocessing steps may often introduce a systematic bias in the image data. Therefore, the rest of the analysis in this paper has been shown only with the raw image data with an aim of showing class discrimination and not using any transformed domain image data.

### *3.3. Fractal Analysis and Hausdorff Dimension*

For fractal dimension calculation, the grayscale images need to be converted to a binary image for the box-counting based fractal analysis. Setting an optimum threshold has always been a crucial step in fractal analysis since an arbitrarily chosen threshold may spuriously increase or decrease the ratio of black and white pixels and change the pattern of the original continuous grayscale image in its discretised binary version. For example, a threshold at the centre of the graylevel (128), may result in disproportionate white and black pixels, with arbitrary modification of the group mean and variance values, thus distorting the fine granularity of the original grayscale image in its binary version. We here use the Otsu's method [95] for automatic determination of the threshold which minimizes the intra-class variance between the black and white pixels.

The box-counting algorithm returns the number of black boxes ($N_b$) of size $R$. For faster numerical computation the box sizes are increased as powers of 2 i.e. $R = 2^0, 2^1, 2^2, \cdots, 2^P$ number of pixels. The algorithm terminates when the maximum of size of the image in either $x$ or $y$ direction becomes less than $2^P$. The negative slope of the box-size ($R$) vs. box-counting ($N_b$) curve in log-log plot, is known as the fractal dimension or Hausdorff dimension ($D_f$), and their empirical power law scaling is given by (10).

$$N_b \sim R^{-D_f} \tag{10}$$

In order to find out the fractal dimension $D_f$, the natural logarithm of both box-size ($\log R$) and box-counting ($\log N_b$) were computed first, followed by a least-square regression with a first order polynomial (straight line) fitting [96]. The negative slope of this line now indicates the fractal dimension for four different cases of biofilms on the HDPE substrate in Figure 3, using either a bare surface, or with a conditioning layer of BSA adsorbed for 9 hours/24 hours or with a conditioning layer of FN 24 hours.

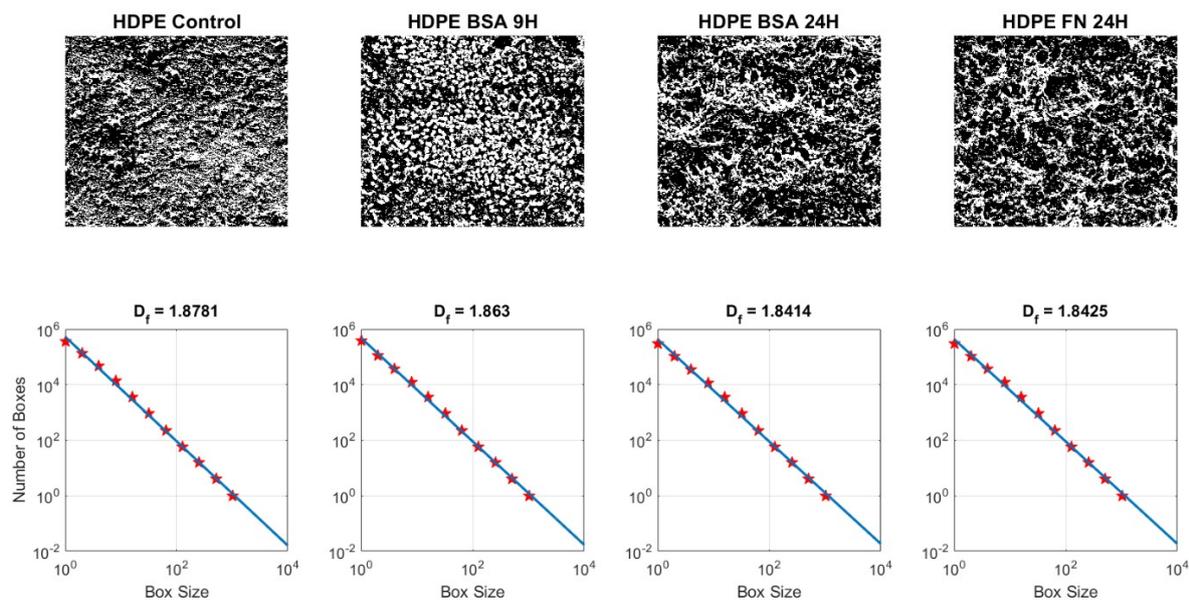

*Figure 3: (Top panel) Binary image after thresholding the grayscale image for HDPE substrate, (Bottom panel) Box-counting fractal dimension estimation of the respective cases.*

A similar exploration on the PTFE substrate has been shown in Figure 4. It is also important to notice that there is minimal loss of basic image characteristics using the optimal image thresholding method to convert the original grayscale SEM image in a binary image for fractal analysis. A different image threshold, if not chosen optimally, could affect the box-counting and hence the fractal dimension estimates. However this has not been explored in the present analysis, as it would be a digression from the main topic of this paper.

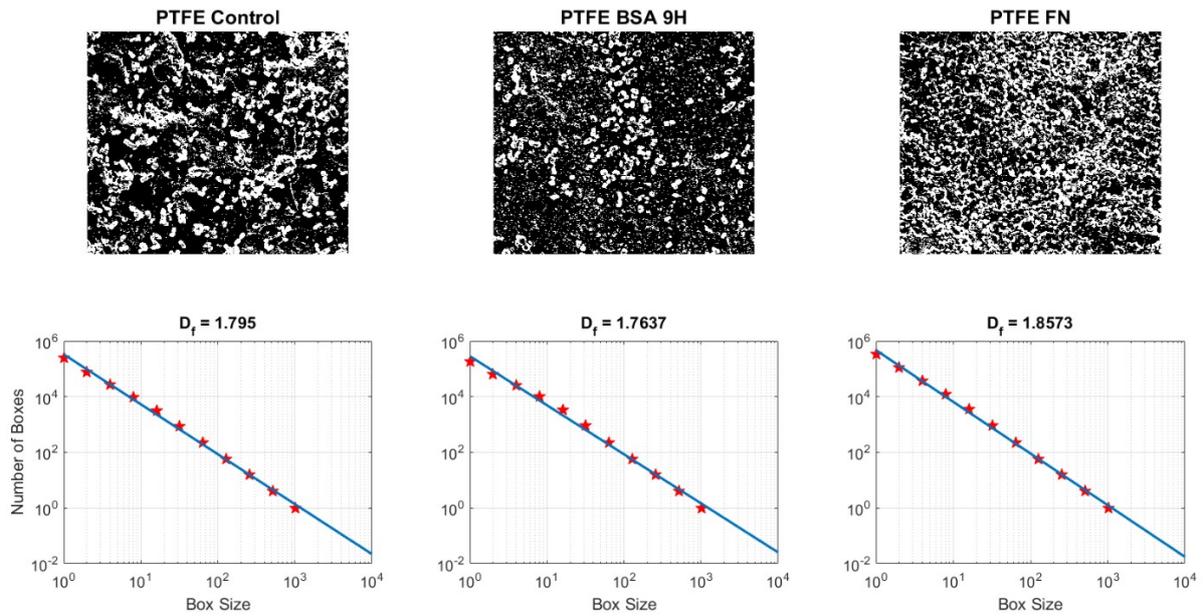

*Figure 4: (Top panel) Binary image after thresholding the grayscale image for PTFE substrate, (Bottom panel) Box-counting fractal dimension estimation of the corresponding binary images.*

### *3.4. Effect of Contrast Enhancement as Preprocessing*

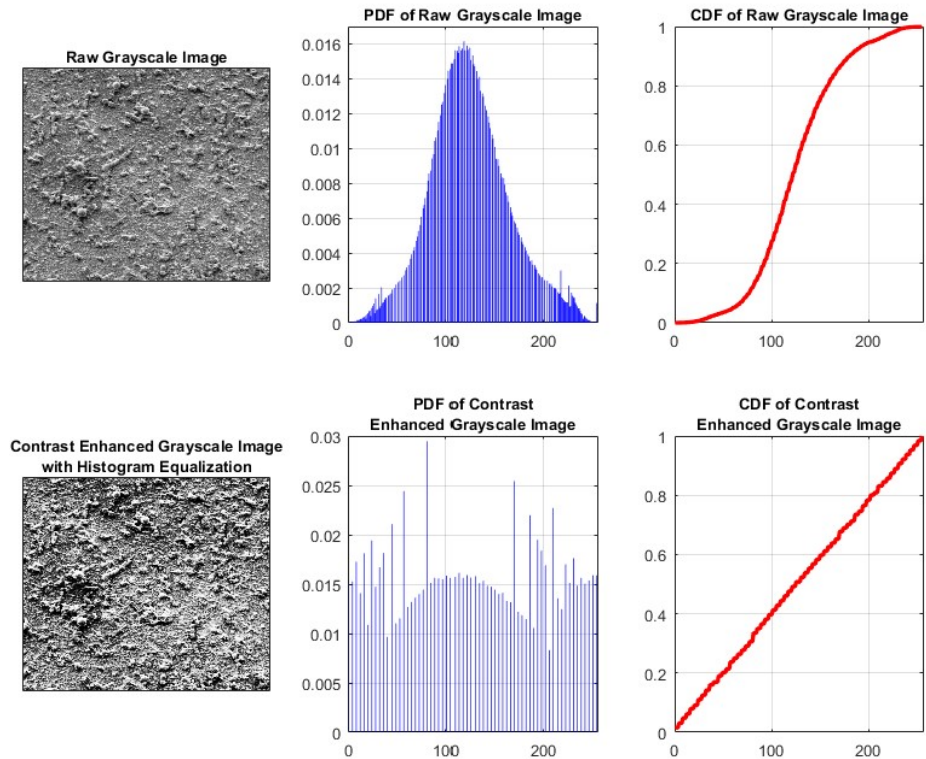

*Figure 5: Effect of contrast enhancement on the image pdf and cdf.*

The morphological features discussed in the previous sections could have been extracted after some preprocessing of the raw image, e.g. different kind of filtering to improve image clarity. One such method know as contrast enhancement via histogram equalization, is widely used in the image processing literature before feature exraction. An example of this procedure is shown in Figure 5, where the raw grayscale image shows a sigmoid type cumulative distribution function (CDF) of the pixel intensity. The purpose of the contrast enhancement is to make the PDF of the grayscale image close to a uniform distribution, so that the corresponding CDF becomes a ramp like function, with a steady increase in pixel intensity over all the bins. In general the clarity of images can be increased with contrast enhancement and thus may often help in textural feature extraction. In this study, we explore its effect on fractal analysis, as the contrast enhancement may drastically change the proportion of black and white pixels in the binary image after thresholding.

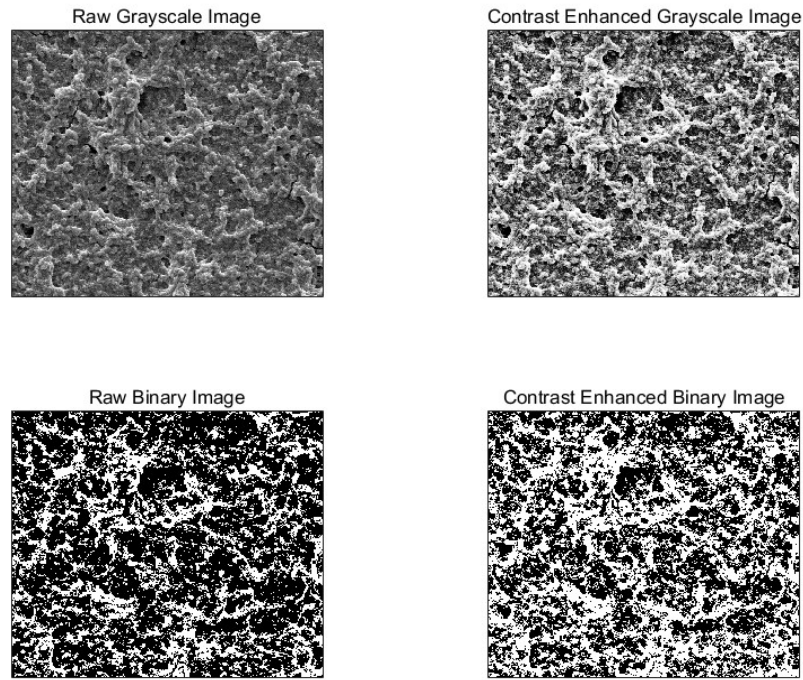

*Figure 6: Effect of contrast enhancement of the raw image in grayscale to binary conversion.*

The effect of contrast enhancement on the image thresholding (for fractal analysis) has been shown in Figure 6. It is evident that although with Otsu's method of automatic thresholding, the contrast enhanced grayscale version produces better clarity of the image, some of the fine structures in the biofilm architecture get smeared in the binary version. A simple binary image however preserves this information. It is also found that with the contrast enhanced version, the same set of features produces higher within group variance compared to the raw image based features. Therefore in the remainder of the statistical analysis, only the raw grayscale images and its binary version have been focussed upon.

### 3.5. Feature Correlation Analysis

Due to lack of independent experiments in our present research and difficulty in producing a number of exactly similar sets of bacterial cultures using the wild clinical *Pseudomonas aeruginosa* strains, the same large images have been resampled with smaller time window, to carry out statistical analysis. This is a viable option for large data where a smaller segment of the data can be assumed to come from different experiments. All the SEM images chosen for the present analysis have a magnification factor 10,000x. For calculating the above mentioned first/second order statistical features and fractal dimension, a smaller blocks of 512×512 pixels have been considered as a fixed scanning window for each of the original images which was slid by 10 pixels along both the *x* and *y* directions for calculating resampled statistics from the original SEM images. This method scans through the different local properties of an image and check the consistency of the features using the criteria of smaller within class variance.

In each of the groups under study, at least 2 sets of independent SEM images were taken and in some cases 3 sets, for a single experimental condition at the same magnification factor. Since biofilm formation is a stochastic process, it is found to be absolutely nonuniform in stray portions of the substrate while it has a definite pattern on rest of the substrate. The SEM images which were not of good quality and having absolutely non-repetitive pattern were outright discarded. Also the SEM images having varying magnification factor were not included in the present study. Hence finally only those SEM images which passed this criteria of good quality and repetitive patterns were included in the study. In the present study, it was difficult to repeat many experiments with clinical or wild bacterial strains as the same strain may not be available repetitively. The protocol for biofilm formation however remanins the same throughout our experiments for all substrates and all conditioning layers for all strains of the microbes. Hence the experimental results can be considered to be quite robust as also reported previously in Dutta Sinha *et al.* [89]. The calculation of different statistical features were done on each set of data separately (also by resampling each image) and their variability has also been reported in the final results.

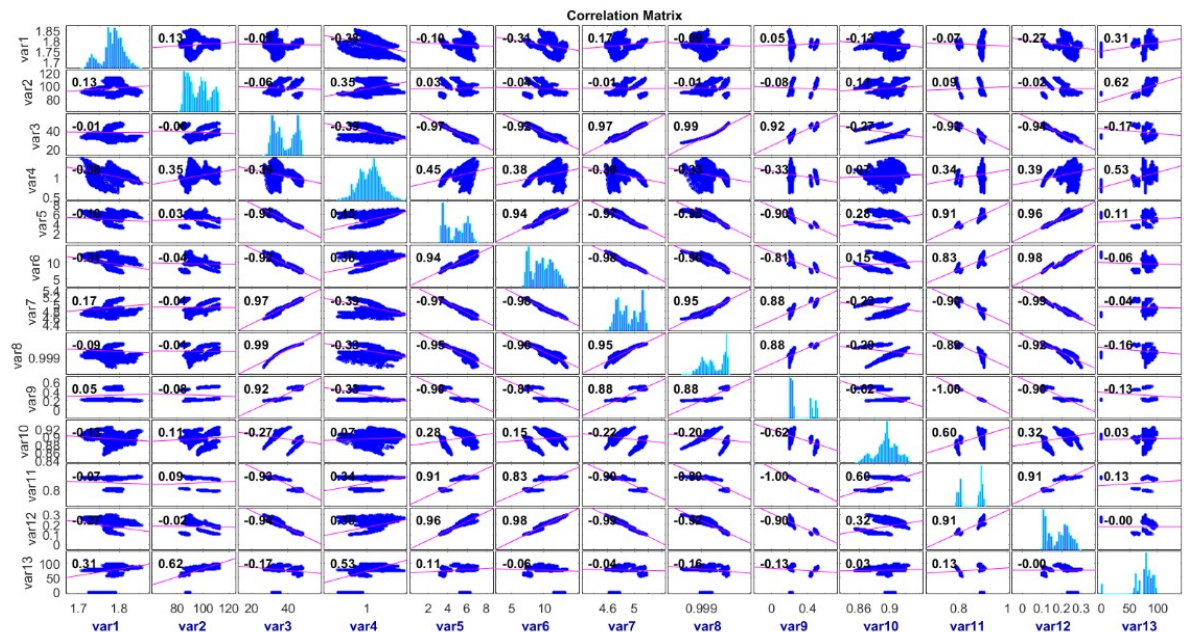

*Figure 7: PTFE control group feature correlation plot.*

In statistical feature based grouping analysis, it is often found that some of the features may be closely related or correlated. In order to verify it, we here carry out a feature correlation analysis as explored in Figure 7 on the PTFE bare surface, for benchmarking purposes. The features or variables are taken in the following order, before using any ranking for group analysis – fractal dimension (var1), mean (var2), standard deviation (var3), skewness (var4), kurtosis (var5), energy (var6), entropy (var7), smoothness (var8), contrast (var9), correlation (var10), homogeneity (var11), 2D energy (var12), mode (var13). In some cases of the joint distributions, a high (>0.9) value of the correlation coefficient between a pair of features is observed. However a conclusive statement of the correlation analysis needs to be verified on a large dataset. On the PTFE substrate (as shown in Figure 7), the scatter plots show much wider variance and less correlation, indicating new information about the structures of the biofilm SEM images captured by most of the features. A variable correlation analysis helps to identify the redundant information going into the class discrimination problem. In the next subsection these features are first ranked according to their class-separability measures, and then used for hypothesis testing.

### 3.6. Class Separability Measures and Feature Ranking

For binary class problem Fisher's discriminant ratio (FDR) is a widely used measure for ranking features according to their statistical discrimination ability. When dealing with a multi-class discrimination problem, usually the scatter matrix in eqn. (11) is used to quantify the class separability using one/multiple features but the effectiveness of each individual feature needs to be judged at the outset [97]. The scatter matrix is given by equation (11), as an extension of FDR for multiclass statistical discrimination problem.

$$J = trace\{S_w^{-1} S_b\}$$
$$S_w = \sum_{i=1}^{c} P_i S_i,$$
$$S_b = \sum_{i=1}^{c} P_i (m_i - m_0)(m_i - m_0)^T, m_0 = \sum_{i=1}^{c} P_i m_i$$

(11)

Here, $\{S_m, S_w, S_b\}$ represent mixture scatter matrix, within-class scatter matrix and between-class scatter matrix respectively. $P_i$ is the *a priori* probability of each class $i$ with respective mean and covariance being $\{m_i, S_i\}$. Also, $m_0$ is known as the global mean vector that considers data points from all the classes.

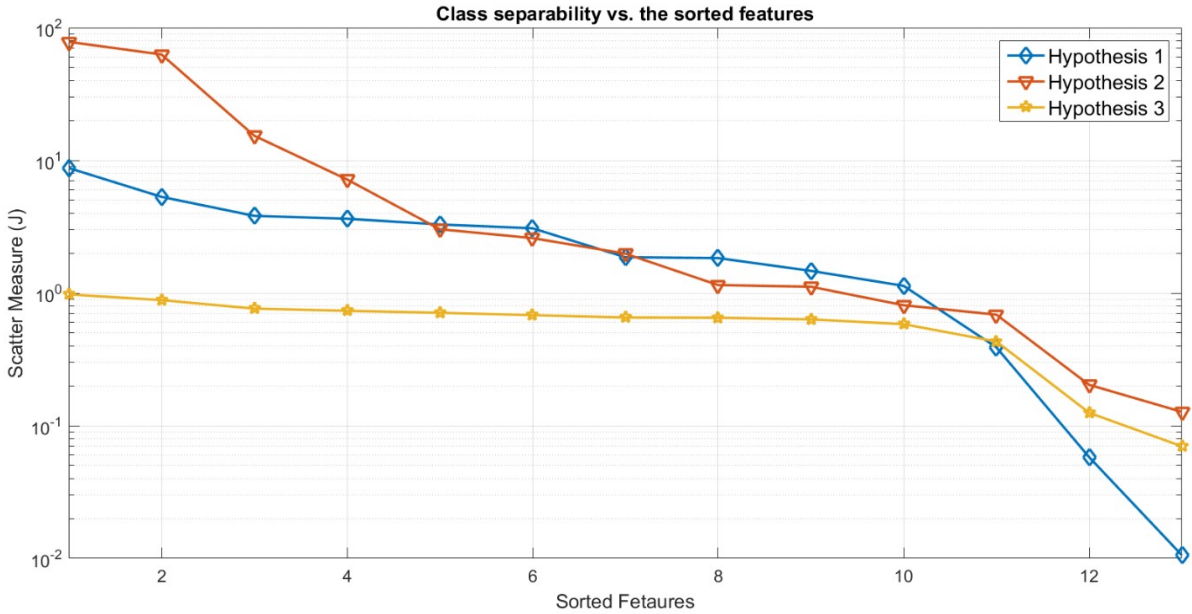

*Figure 8: Variation in class separability measure based on scatter matrix with sorted features in the three discrimination problems. The features are sorted using decreasing value of scatter measure J.*

Considering each feature individually the above covariance matrices reduce to the individual variance. We here rank the features using the univariate scatter score $J$. The scatter histograms of the top three features have also been explored later on, to show the class separability performance in a 2-dimensional feature space. Before calculating the scatter measures, each feature (*f*) has been standardized to zero mean and unit variance to avoid any bias due to different ranges of features using

$$f_{std}^i = \left(f^i - \mu_f^i\right)/\sigma_f^i, \qquad (12)$$

where $\{\mu_f^i, \sigma_f^i\}$ denote the mean and variance of $i^{th}$ feature respectively.

According to the three distinct hypotheses mentioned in section 2.4, the above 13 statistical features are ranked first according to their decreasing level of significance and shown in Figure 8. It is understandable that for each of these hypothesis all the features will not get a similar priority which is explored in the next section.

### 3.7. Hypothesis Testing

Usually the one way analysis of variance (ANOVA) tests the null hypothesis that the mean values of multiple groups are different with an underlying assumption that all the samples in each group came a Gaussian distribution with a common variance. In more realistic cases, this condition is violated and a non-parametric version of one way ANOVA i.e. the Kruskal-Wallis test is preferred [98]. Also the classical *F*-statistic in ANOVA is replaced by $\chi^2$ statistic and the significance level is measured by the *p*-value. A result is considered to be significant if the *p*-value is below a certain value. The null hypothesis here for the non-parametric test is that the features from different groups came from the same distribution and their median values are the same, against the alternative hypothesis that their medians are different. A low *p*-value rejects the null hypothesis and the alternative hypothesis is accepted. Due to multiple comparisons on the same data (although using different features), the significance levels are corrected using Bonferroni's method by dividing the actual *p*-values ($p<0.01$) with the number of tests conducted on the same data.

## 4. Results and Discussions
### 4.1. Results for Hypothesis 1: Effect of Adsorption Time Variation on HDPE

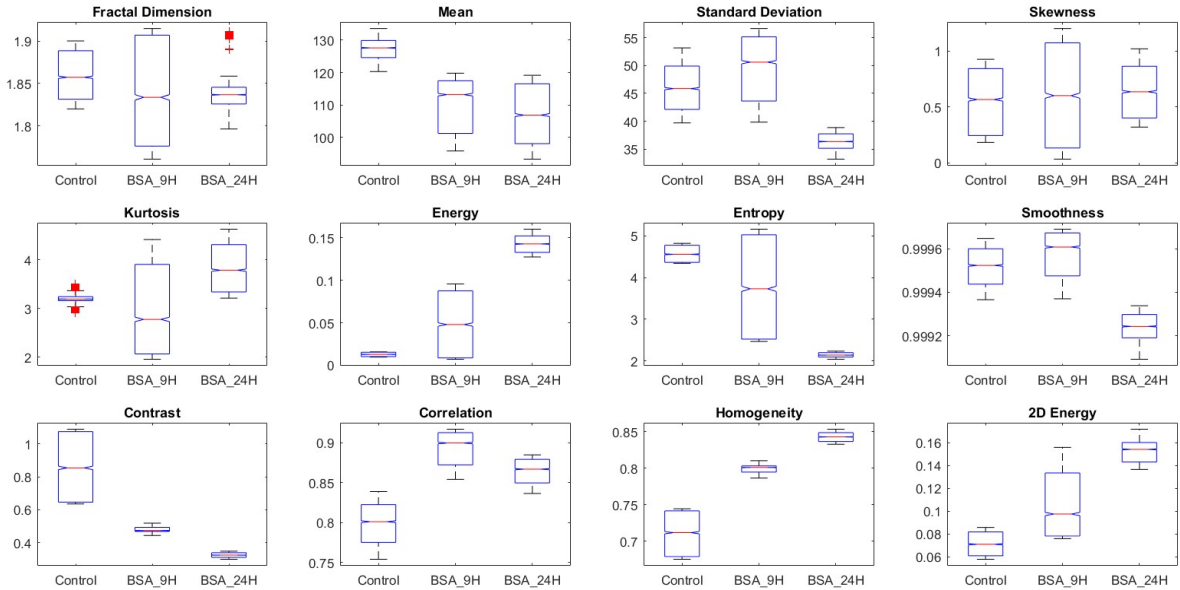

*Figure 9: Feature box plot for BSA conditioning layer on HDPE substrate for 9H vs. 24H*

The ranked features for the hypothesis 1 capturing the effect of protein adsorption time on HDPE substrate has been listed in Table 1, along with the significance levels of the Kruskal-Wallis test on each feature with Bonferroni correction. The features are ranked using the scatter matrix (*J*) showing a clear separation of the central tendency (mean/median) and small within class variance. The $\chi^2$ statistic corresponding to each ANOVA table has been reported showing a gradual decrease in $\chi^2$ statistic with the scatter matrix (*J*). The results have been graphically visualised in the form of boxplots for 12 features (except the mode) in Figure 9. In all the boxplots, the median, interquartile range (IQR) and outliers are shown using the red horizontal line, blue box and red cross markers respectively. Here, the fractal dimension for the HDPE BSA are found to have a minimum within class variance and ranging between 1.8-1.85. The higher moments like skewness and kurtosis also shows the deviation from the Gaussian behaviour justifying the choice of nonparametric hypothesis test over the parametric ANOVA test. The steady increase in energy is almost found to have an inverse relation with a steady decrease in entropy when the substrate for biofilm growth changed from HDPE bare surface to HDPE surface adsorbed with BSA for 9 hours and 24 hours. The contrast, homogeneity and 2D energy also show either an increasing or decreasing pattern in these three conditions.

*Table 1: Sorted features for BSA conditioning layer on HDPE substrate for 9H vs. 24H based on decreasing J*

| *Scatter Measure J* | **Feature Description** | $\chi^2$ | *p*-value |
|---|---|---|---|
| 8.761 | $F_1$ = Homogeneity | $1.054 \times 10^4$ | 0 |
| 5.311 | $F_2$ = Energy | $7.903 \times 10^3$ | 0 |
| 3.819 | $F_3$ = Correlation | $9.048 \times 10^3$ | 0 |
| 3.632 | $F_4$ = 2D Energy | $8.439 \times 10^3$ | 0 |
| 3.288 | $F_5$ = Contrast | $1.054 \times 10^4$ | 0 |
| 3.078 | $F_6$ = Smoothness | $8.348 \times 10^3$ | 0 |
| 1.863 | $F_7$ = Entropy | $7.903 \times 10^3$ | 0 |
| 1.835 | $F_8$ = Standard Deviation | $8.348 \times 10^3$ | 0 |
| 1.470 | $F_9$ = Mean | $8.067 \times 10^3$ | 0 |
| 1.133 | $F_{10}$ = Mode | $7.421 \times 10^3$ | 0 |
| 0.3896 | $F_{11}$ = Kurtosis | $4.003 \times 10^3$ | 0 |
| 0.05793 | $F_{12}$ = Fractal Dimension | $3.943 \times 10^2$ | $2.417 \times 10^{-86}$ |
| 0.01055 | $F_{13}$ = Skewness | $3.277 \times 10^2$ | $6.838 \times 10^{-72}$ |

The results can also be viewed as scatter plots or joint distributions of the most significant set of features. The clear separation of the three groups amongst different combinations of energy, homogeneity and correlation are explored in Figure 10. A kernel density smoothing is used to compute the marginal distributions (projections on either *X* or *Y* axes) from the scatter diagrams of the feature pairs of the three groups. Although in some cases the 1D marginal distributions show some degree of overlap between different groups, in a higher dimensional space, i.e. considering feature combinations, instead of simple class discrimination using hypothesis tests reported above will result in a very clear separation between these groups. The gradual separation of the three groups in 1D box plot to 2D and 3D scatter plot indicates that for such class separation problem, feature pairs or groups should definitely be used to quantify and discriminate different biofilm morphologies.

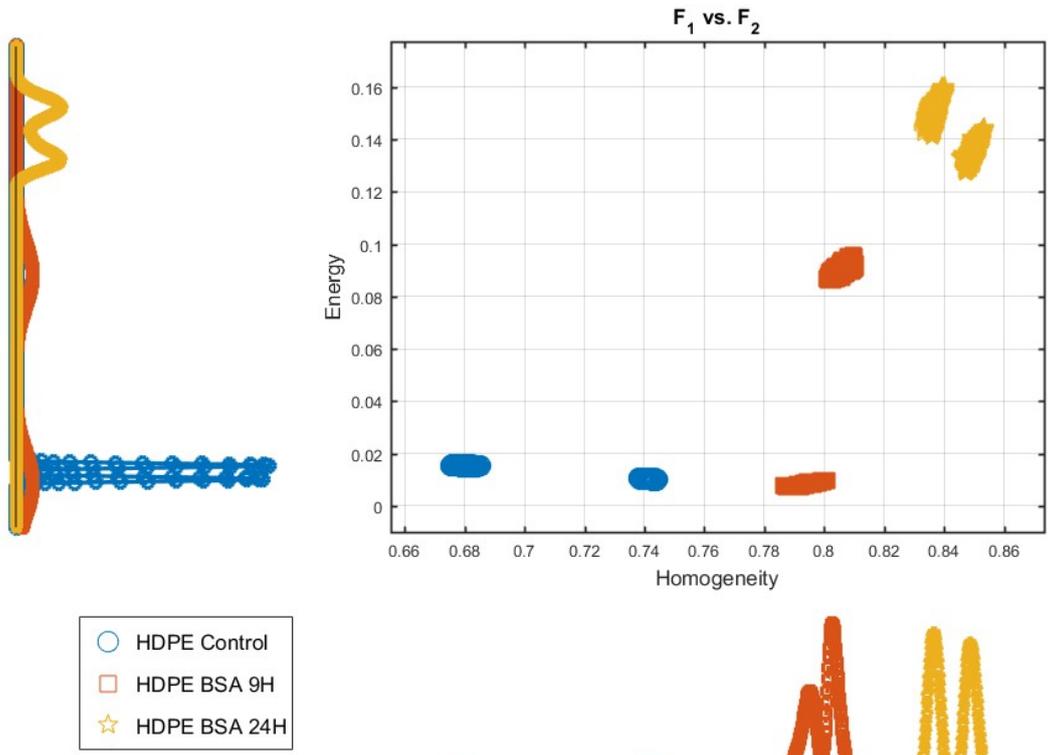

*(a)*

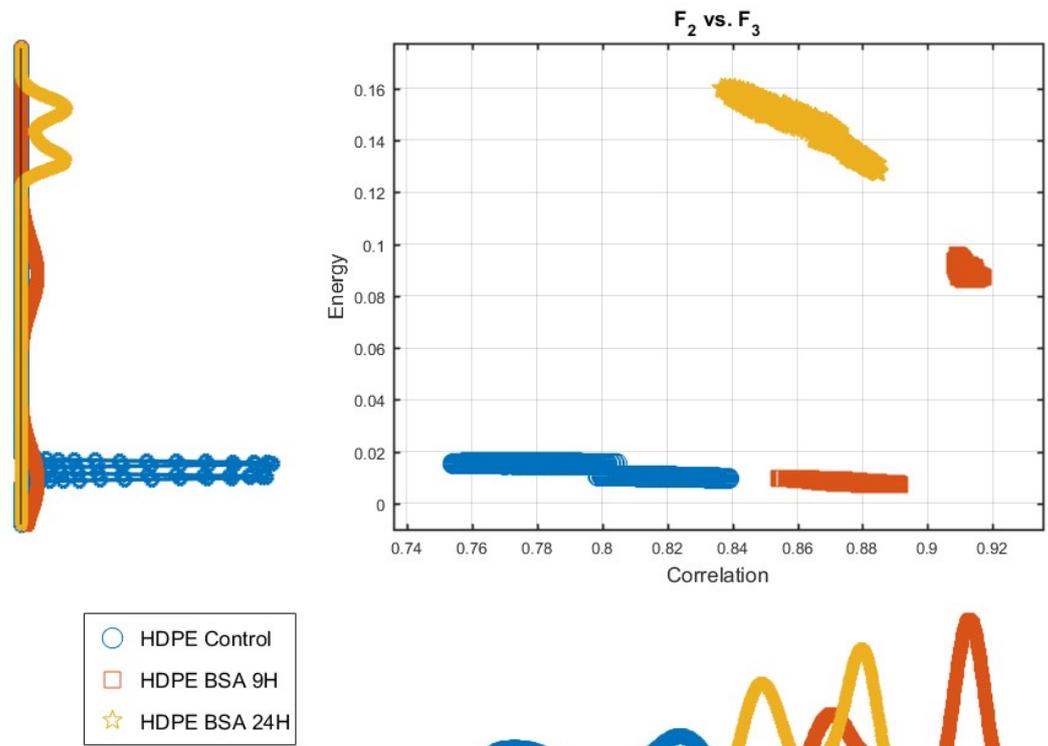

*(b)*

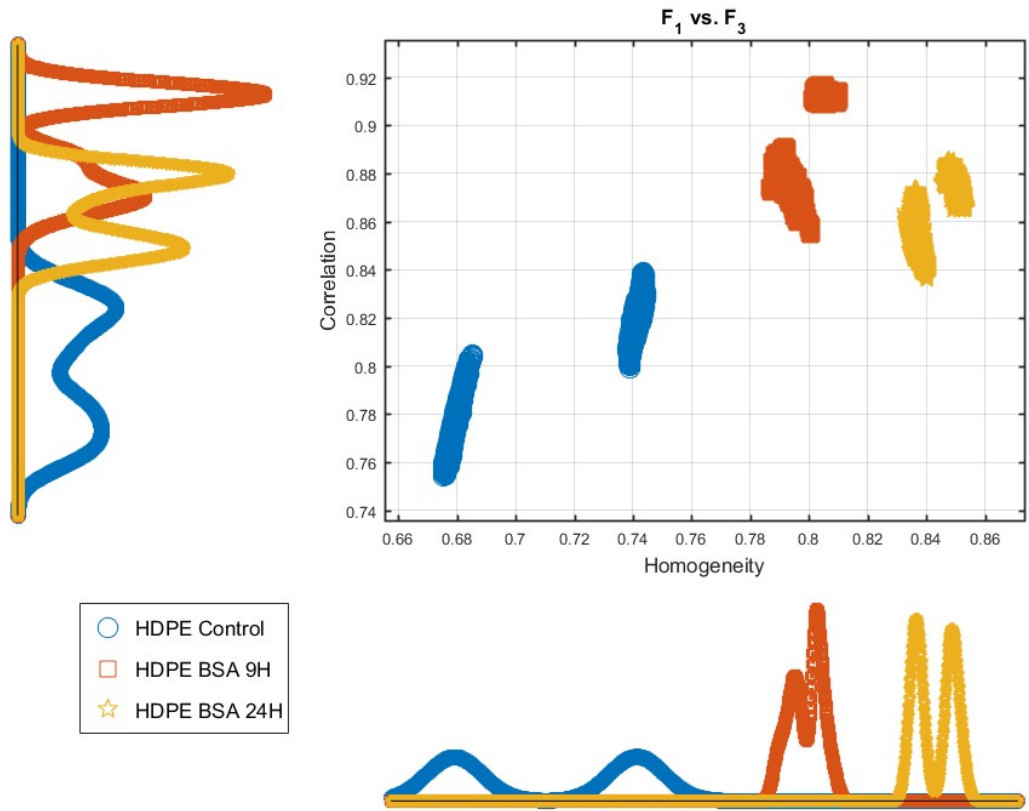

*(c)*

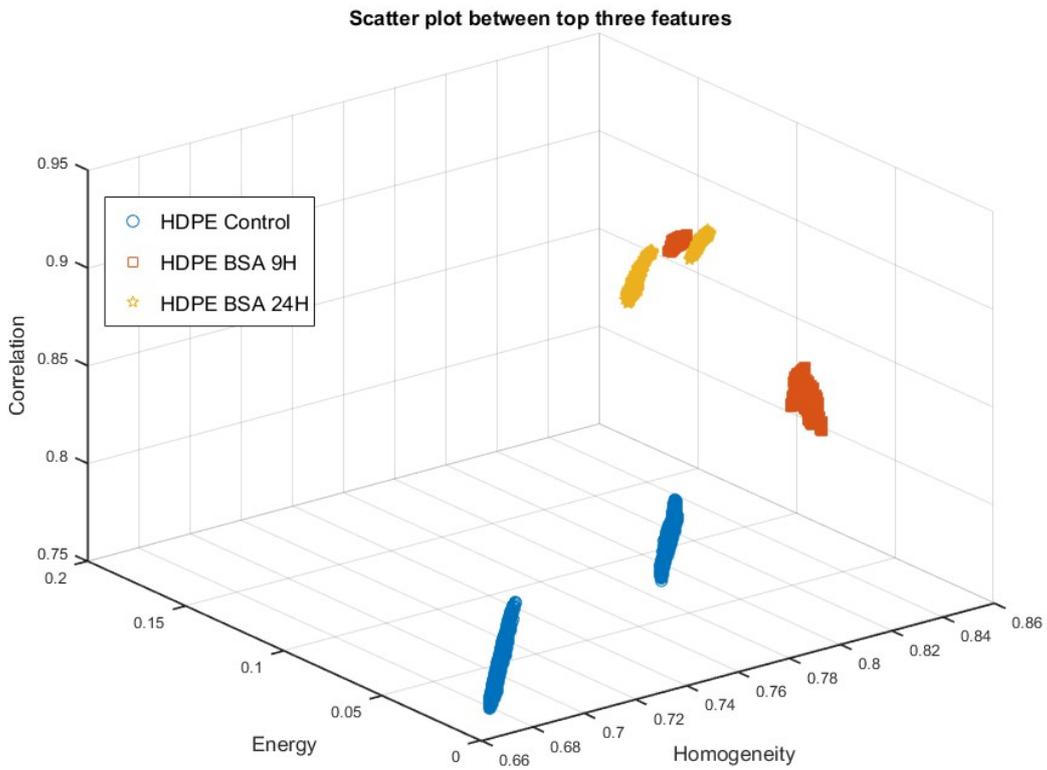

*(d)*

*Figure 10: Scatter plots and histograms of top three features for hypothesis 1: effect of adsorption time on HDPE (a) $F_1$ vs. $F_2$, (b) $F_2$ vs. $F_3$, (c) $F_1$ vs. $F_3$, (d) $F_1$-$F_3$*

## 4.2. Results for Hypothesis 2: Effect of Variation of Conditioning Layer BSA vs. FN

A similar statistical analysis has been carried out to test the effect of different proteins (BSA and FN) on the same HDPE substrate, using these features as shown in Table 2. The fractal dimension in this case has been found to be more compact for the HDPE BSA 24 hours case, as shown in Figure 11. Apparently from the scatter diagram of energy and entropy in Figure 12(a), the marginal distributions may appear to be overlapping. However the clear separation between these two groups using another feature 2D energy can be viewed in the Figure 12(b)-(d) in the 2D and 3D feature space.

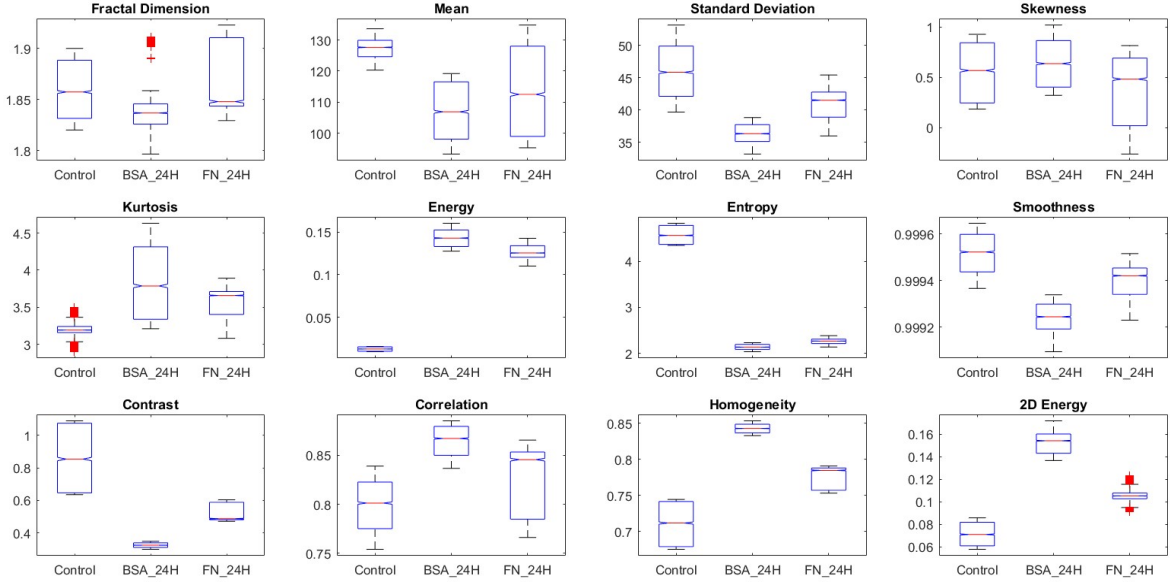

*Figure 11: Feature box plot for HDPE substrate with different conditioning layers BSA 24H vs. FN 24H.*

*Table 2: Sorted features for HDPE substrate with different conditioning layers BSA 24H vs. FN 24H based on decreasing J*

| *Scatter Measure J* | **Feature Description** | $\chi^2$ | *p*-value |
|---|---|---|---|
| 78.25 | $F_1$ = Entropy | $7.903 \times 10^3$ | 0 |
| 62.86 | $F_2$ = Energy | $7.903 \times 10^3$ | 0 |
| 15.34 | $F_3$ = 2D Energy | $8.439 \times 10^3$ | 0 |
| 7.182 | $F_4$ = Homogeneity | $1.054 \times 10^4$ | 0 |
| 3.028 | $F_5$ = Contrast | $1.054 \times 10^4$ | 0 |
| 2.595 | $F_6$ = Smoothness | $8.348 \times 10^3$ | 0 |
| 1.982 | $F_7$ = Standard Deviation | $8.348 \times 10^3$ | 0 |
| 1.149 | $F_8$ = Correlation | $9.048 \times 10^3$ | 0 |
| 1.116 | $F_9$ = Mode | $7.421 \times 10^3$ | 0 |
| 0.8098 | $F_{10}$ = Mean | $8.067 \times 10^3$ | 0 |
| 0.687 | $F_{11}$ = Kurtosis | $4.003 \times 10^3$ | 0 |
| 0.2033 | $F_{12}$ = Fractal Dimension | $3.943 \times 10^2$ | $2.417 \times 10^{-86}$ |
| 0.1272 | $F_{13}$ = Skewness | $3.277 \times 10^2$ | $6.838 \times 10^{-72}$ |

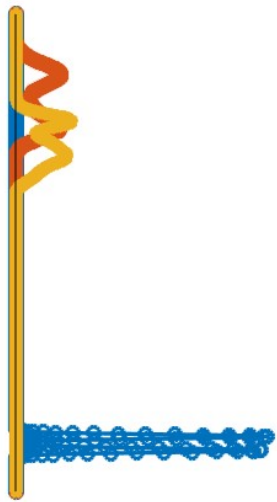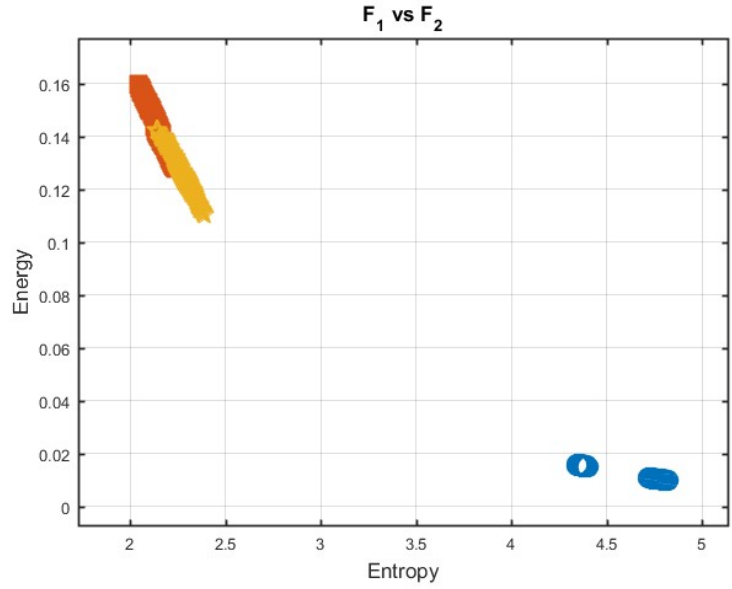
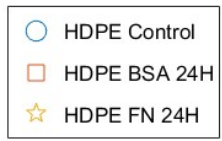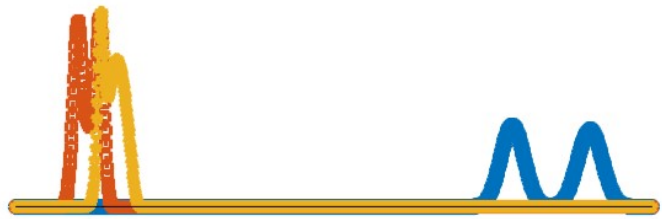

*(a)*

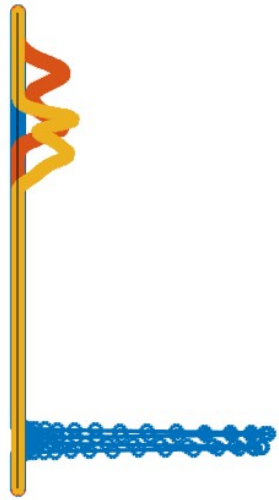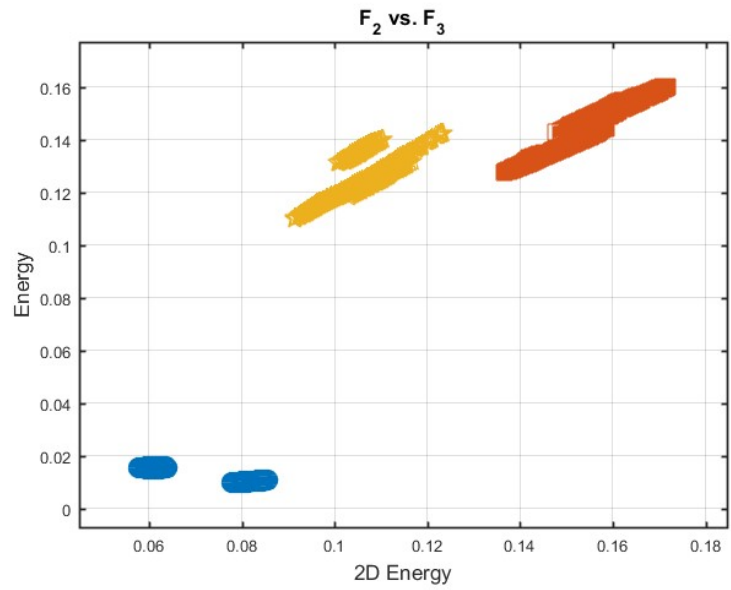
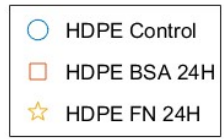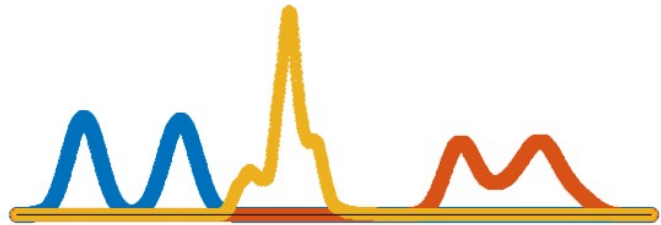

*(b)*

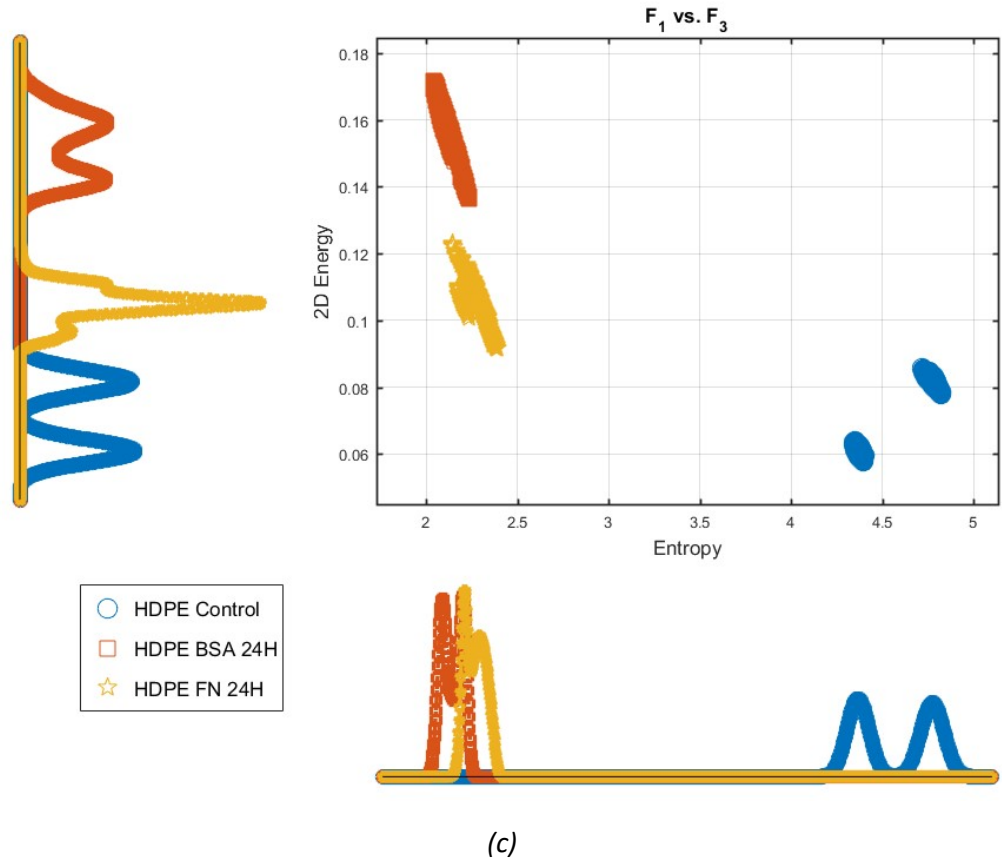

*(c)*

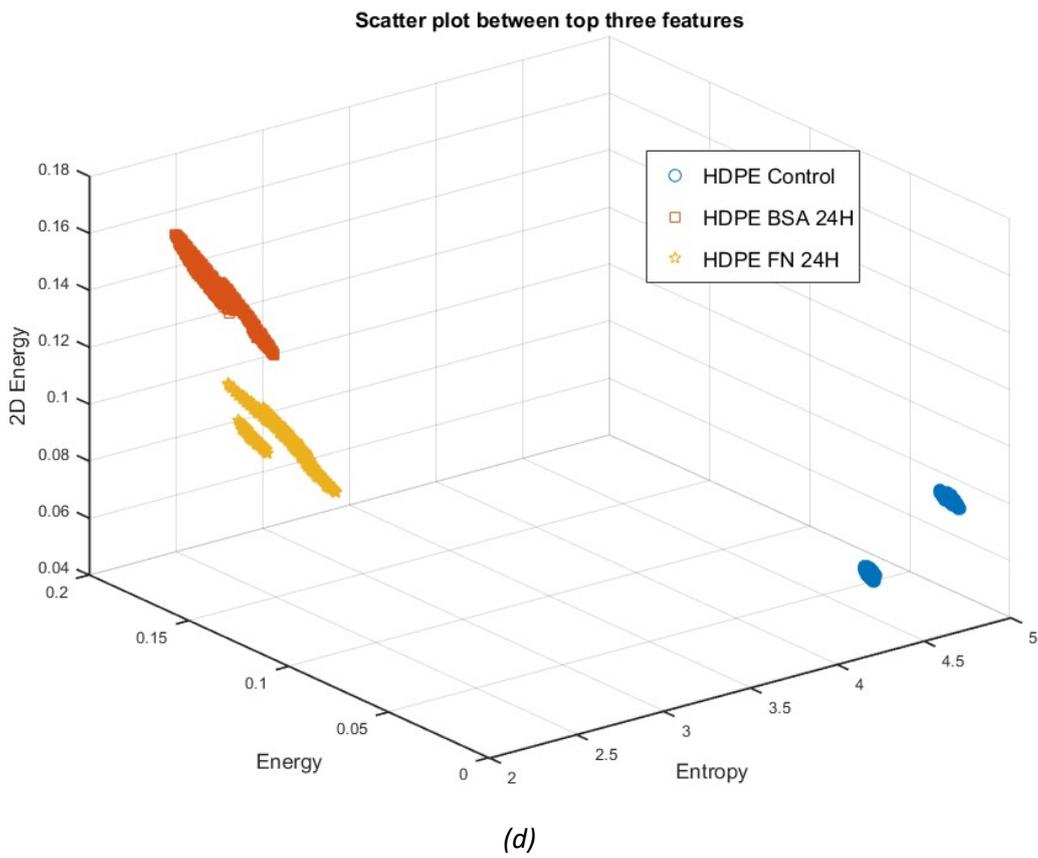

*(d)*

*Figure 12: Scatter plots and histograms of top three features for hypothesis 2: effect of conditioning layer BSA vs. FN on HDPE substrate (a) $F_1$ vs. $F_2$, (b) $F_2$ vs. $F_3$, (c) $F_1$ vs. $F_3$, (d) $F_1$-$F_3$*

## 4.3. Results for Hypothesis 3: Testing on another Substrate PTFE

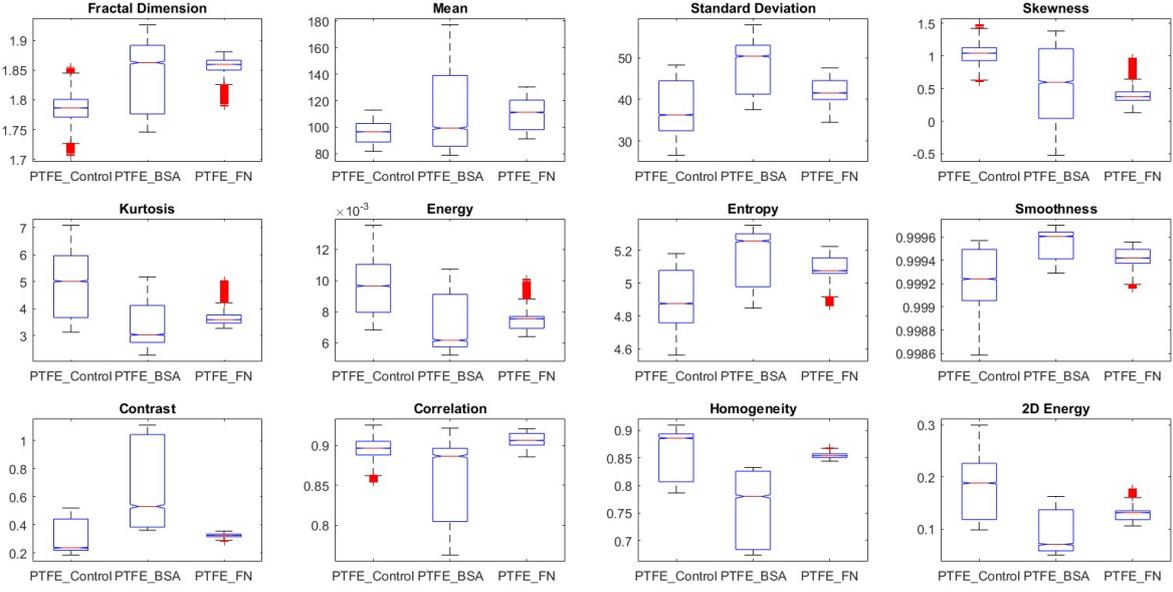

*Figure 13: Feature box plot for PTFE substrate with different conditioning layers BSA 24H vs. FN 24H.*

*Table 3: Sorted features for PTFE substrate with different conditioning layers BSA 24H vs. FN 24H based on decreasing J*

| Scatter Measure J | Feature Description | $\chi^2$ | $p$-value |
|---|---|---|---|
| 0.9764 | $F_1$ = Homogeneity | $1.054 \times 10^4$ | 0 |
| 0.8848 | $F_2$ = 2D Energy | $8.439 \times 10^3$ | 0 |
| 0.7633 | $F_3$ = Fractal Dimension | $3.943 \times 10^2$ | $2.417 \times 10^{-86}$ |
| 0.7345 | $F_4$ = Contrast | $1.054 \times 10^4$ | 0 |
| 0.7095 | $F_5$ = Kurtosis | $4.003 \times 10^3$ | 0 |
| 0.6814 | $F_6$ = Skewness | $3.277 \times 10^2$ | $6.838 \times 10^{-72}$ |
| 0.6542 | $F_7$ = Standard Deviation | $8.348 \times 10^3$ | 0 |
| 0.6514 | $F_8$ = Entropy | $7.903 \times 10^3$ | 0 |
| 0.6334 | $F_9$ = Energy | $7.903 \times 10^3$ | 0 |
| 0.5828 | $F_{10}$ = Smoothness | $8.348 \times 10^3$ | 0 |
| 0.4313 | $F_{11}$ = Correlation | $9.048 \times 10^3$ | 0 |
| 0.1252 | $F_{12}$ = Mean | $8.067 \times 10^3$ | 0 |
| 0.06988 | $F_{13}$ = Mode | $7.421 \times 10^3$ | 0 |

As the third hypothesis while testing characteristics of biofilm architectures on the PTFE substrate the fractal dimension is found to be more compact with FN protein. The higher standardized moments – skewness and kurtosis show significant deviation from Gaussianity for all the cases. Also the inverse relationship between energy and entropy can still be inspected in the boxplots in Figure 13. The decrease in $\chi^2$ statistic of the ANOVA table is found to be similar to the decreasing scatter matrix in Table 3, as also found before. The 2D scatter plots using the top 3 features in Figure 14 shows many compact but disjoint islands, resulting in some overlapping regions in the 1D marginal distributions, especially using fractal dimension and 2D energy. However the class separation in the 3D feature space is clear in the 3D scatter diagram in Figure 14(d), as different groups form non-overlapping disjoint clouds.

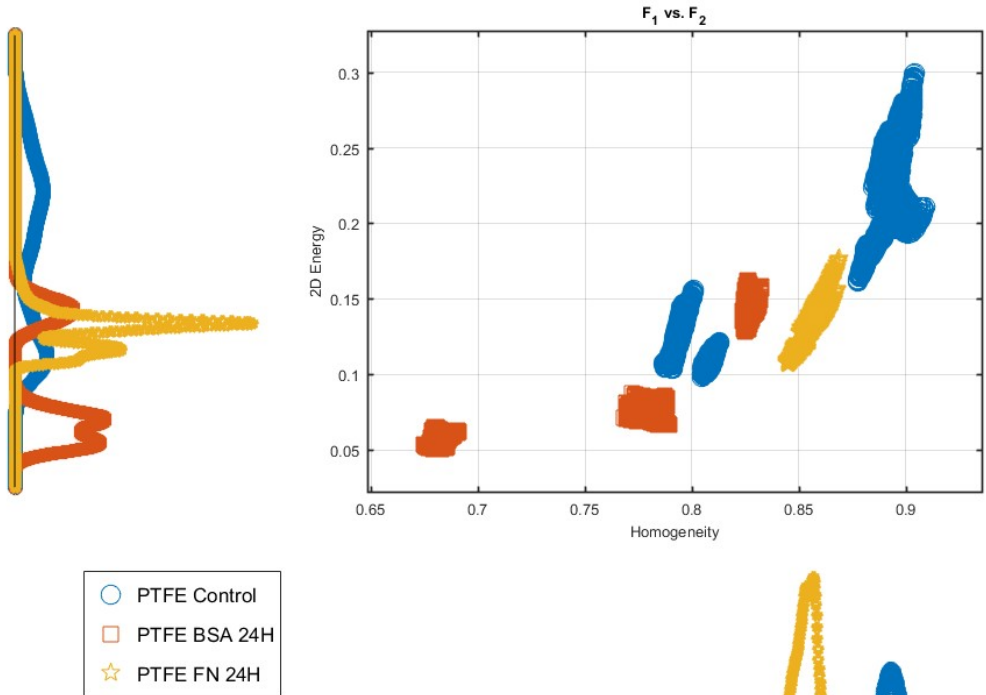

*(a)*

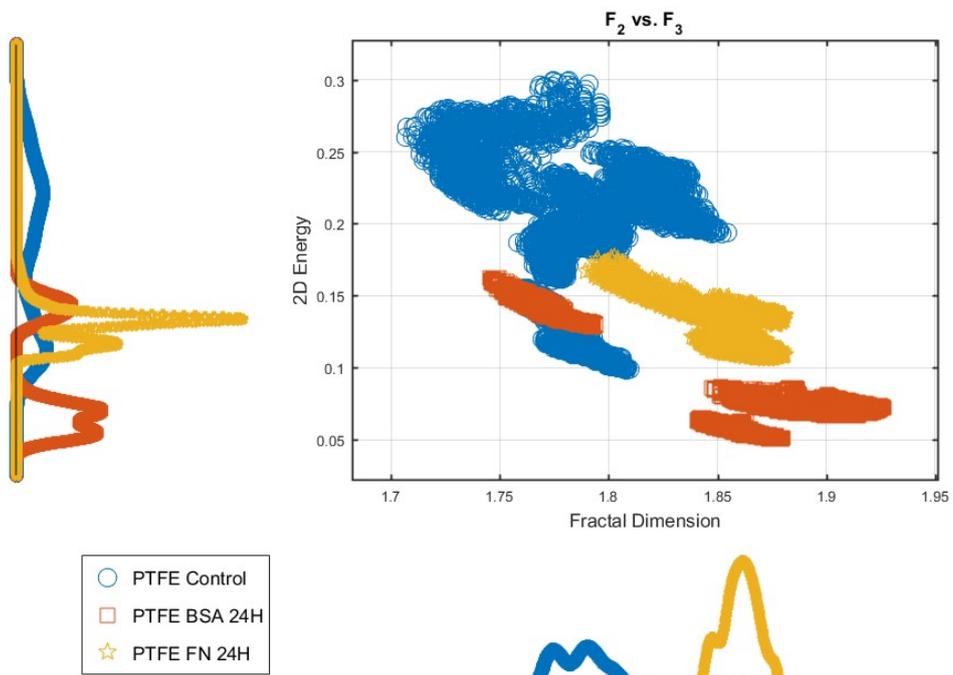

*(b)*

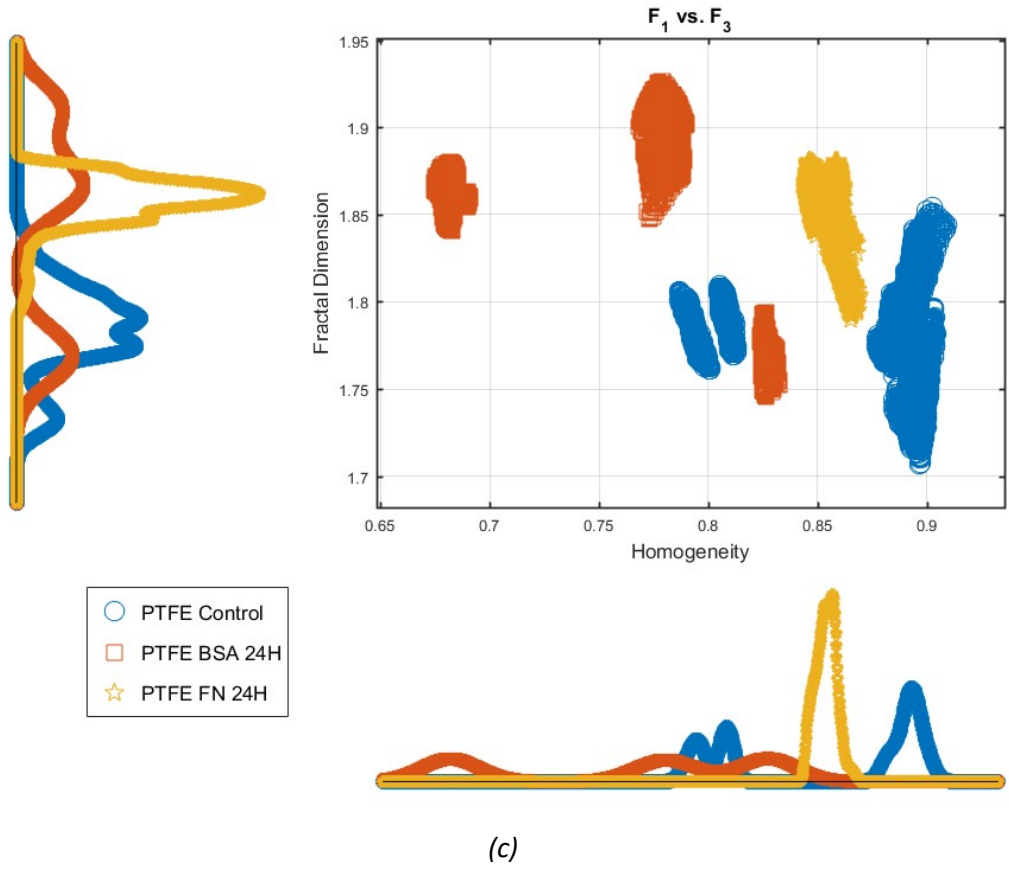

*(c)*

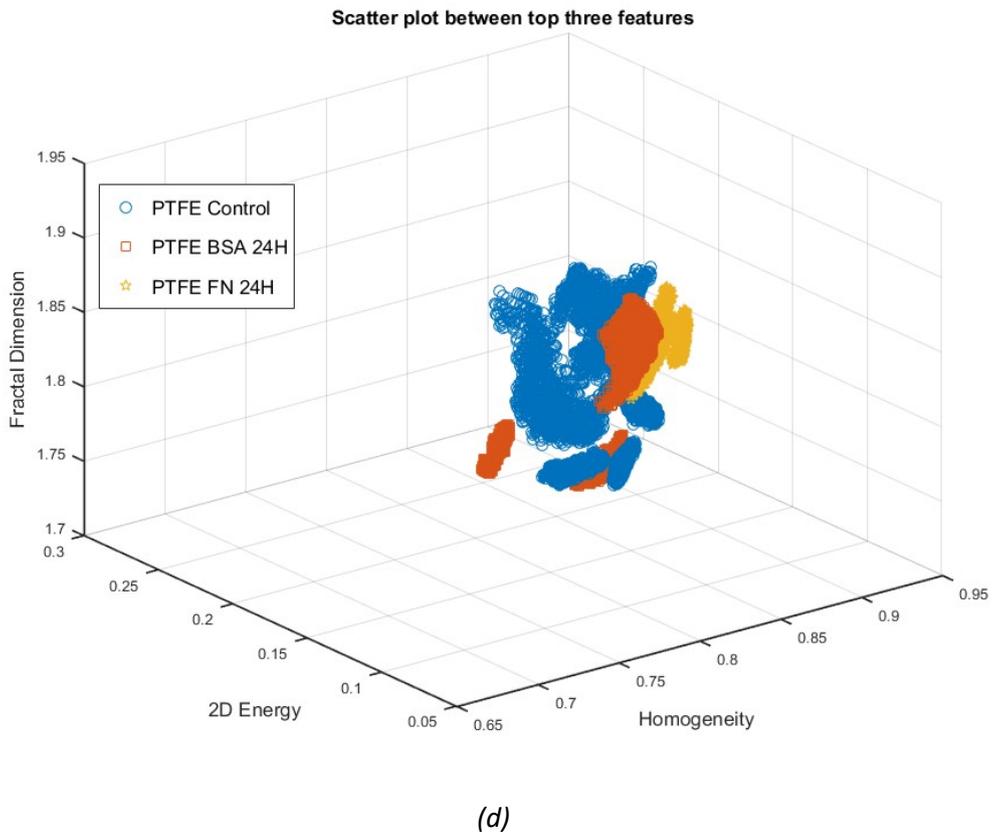

*(d)*

*Figure 14: Scatter plots and histograms of top three features for hypothesis 3: effect of conditioning layer BSA vs. FN on PTFE substrate (a) $F_1$ vs. $F_2$, (b) $F_2$ vs. $F_3$, (c) $F_1$ vs. $F_3$, (d) $F_1$-$F_3$*

*4.4. Discussions*

It is observed from the above reults that different features emerge for different hypothesis or experimental conditions and the variation of the features (extracted from multiple images taken at the same condition) are reported in the box-plots and already been penalised in the feature selection process using Scatter matrix. For any feature, if the variance is high for same experimental condition (coming from different images) or due to resampling (smaller sub-segments of the same image), both would show a low scatter measure as in Figure 8 due to increased variance if either of the case occur. There is less chance that due to low variance (within and between experiments) and high scatter measure, the top 3 features (as shown in this section) would capture non-repeatable patterns. However to have a conclusive answer on this, we plan to report large scale clinical trials with our collaborative hospitals in a future study. However we feel that repeatability through 2-3 independent tests and resampling on each SEM image to create multiples is an established statistical characterisation method which is adopted here as well.

Since biofilm structures are complex and heterogeneous, there have been lot of effort to analyse such microstructures, but the exact objective of investigation may be different. While some may determine, the amount of EPS matrix, others may quantify the total number of bacterial cells embedded in biofilm or the effective number of living bacteria in biofilm. Such different targets usually require different approaches. Amonsgt many approaches the colorimetric methods [99,100] are capable of quantifying living bacteria in a biofilm. However the target of such colorimetric assay is quantification of living bacterial cells in a particular biofilm and not the biofilm architecture as a result of microbe-substrate interaction. The present paper describes a novel method of quantification of differences in biofilm architecture on different substrates to gain insight about the contribution of the substrate in modulating the biofilm forming capabilities of a specific microbe.

Also, we have chosen to work with a clinical strain, instead of working with reference (laboratory) strains. We have gone forth with the clinical pathogenic strains obtained from the urinary catheters of patients suffering from urinary infections at a public health care unit in Kolkata, India. Our goal was to prove experimentally that a pathogenic strain of bacteria, responsible for urinary infections (forming biofilm on silicon rubber) is very much liable to affect orthopaedic implants, if they get the required access. The present paper shows that the substrate-microbe and conditioning layer affecting the biofilm architecture can be successfully quantified from statistical analysis of SEM images. The study reported here is based on smaller set of clinical strains as a proof of concept study. Large scale analysis of wild clinical strains may be studied in a future work by varying the pathogen species and SEM magnification factor. The objective of this paper is to primarily establish the quantification method by feature extraction, ranking and carrying out hypothesis testing to see separation of different biofilm formation conditions, in the feature space.

5. **Conclusions and Scope of Future Work**

The significant feature sets have been identified for discriminating biofilms of a wild *Pseudomonas aeruginosa* strain based on biofilm growth conditions using SEM images *via* image feature extraction and ranking schemes. Three different biofilm growth conditions, including the substrate, conditioning layer protein and absorption time variation which are tested as three distinct hypotheses using the ranked features and the corresponding statistical significance levels have also been reported. Various textural features related to the fractal nature and first/second order statistics of the SEM images in the respective boxplots, kernel density smoothed histograms and 2D/3D joint distributions or scatter plots also qualitatively show the class separability between these image groups. Different set of features have emerged as the most significant ones for comparing different groups including homogeneity, energy, correlation, entropy, 2D energy and fractal dimension etc. However the analysis can be extended with a much larger database and a large pool of features using various other transformed domain features of the SEM images and can be explored in a future research. The present research can be considered as a step forward in that line and briefly reviews the existing contributions on biofilm characterization through SEM images.

The long term goal of our research may be summarized as the estimation of substrate-microbe interactions using a clinical strain of bacteria on different substrates. Such quantification approach may serve as an inexpensive procedure for quantifying substrate-microbe interactions and hence determine the degree of bio-incompatibility of different substrates. In addition this method can be beneficial to scientists from different disciplines, working on biofilms, but lacking the training of microbiological methods. We would like to extend our work in future to Gram positive bacteria strains and on other biomaterial surfaces. Future scope of work may also include the use of hybrid characterization techniques through multiple imaging techniques like CSLM or AFM and assimilation of experimental data with computer simulation of biofilm patterns.

**Acknowledgement**

SDS acknowledges the funding from the Department of Science and Technology (DST), Govt. of India through the Women's Scientist Scheme – A, project no. LS-466/WOS A/2012-2013.

**Table of Content for Graphics**

*Figure 1: (Top panel) Grayscale biofilm imagesfor HDPE, (Bottom panel) Normalised histogram of the pixel intensities.*

*Figure 2: (Top panel) Grayscale biofilm images for PTFE, (bottom panel) Normalised histogram of the pixel intensities.*

*Figure 3: (Top panel) Binary image after thresholding the grayscale image for HDPE substrate, (Bottom panel) Box-counting fractal dimension estimation of the respective cases.*

*Figure 4: (Top panel) Binary image after thresholding the grayscale image for PTFE substrate, (Bottom panel) Box-counting fractal dimension estimation of the corresponding binary images.*

*Figure 5: Effect of contrast enhancement on the image pdf and cdf.*

*Figure 6: Effect of contrast enhancement of the raw image in grayscale to binary conversion.*

*Figure 7: PTFE control group feature correlation plot.*

*Figure 8: Variation in class separability measure based on scatter matrix with sorted features in the three discrimination problems. The features are sorted using decreasing value of scatter measure J.*

*Figure 9: Feature box plot for BSA conditioning layer on HDPE substrate for 9H vs. 24H*

*Figure 10: Scatter plots and histograms of top three features for hypothesis 1: effect of adsorption time on HDPE (a) $F_1$ vs. $F_2$, (b) $F_2$ vs. $F_3$, (c) $F_1$ vs. $F_3$, (d) $F_1$-$F_3$*

*Figure 11: Feature box plot for HDPE substrate with different conditioning layers BSA 24H vs. FN 24H.*

*Figure 12: Scatter plots and histograms of top three features for hypothesis 2: effect of conditioning layer BSA vs. FN on HDPE substrate (a) $F_1$ vs. $F_2$, (b) $F_2$ vs. $F_3$, (c) $F_1$ vs. $F_3$, (d) $F_1$-$F_3$*

*Figure 13: Feature box plot for PTFE substrate with different conditioning layers BSA 24H vs. FN 24H.*

*Figure 14: Scatter plots and histograms of top three features for hypothesis 3: effect of conditioning layer BSA vs. FN on PTFE substrate (a) $F_1$ vs. $F_2$, (b) $F_2$ vs. $F_3$, (c) $F_1$ vs. $F_3$, (d) $F_1$-$F_3$*